\documentclass[twocolumn,showpacs,superscriptaddress,preprintnumbers,amsmath,amssymb,prl,aps]{revtex4}

\usepackage{graphicx}
\usepackage{dcolumn}
\usepackage{bm}
\usepackage{color}

\newcommand{\TeV}{\ensuremath{\mathrm{Te\kern -0.1em V}}}
\newcommand{\TeVc}{\ensuremath{\mathrm{Te\kern -0.1em V\!/}c}}
\newcommand{\TeVcc}{\ensuremath{\mathrm{Te\kern -0.1em V\!/}c^2}}
\newcommand{\GeV}{\ensuremath{\mathrm{Ge\kern -0.1em V}}}
\newcommand{\GeVc}{\ensuremath{\mathrm{Ge\kern -0.1em V\!/}c}}
\newcommand{\GeVcc}{\ensuremath{\mathrm{Ge\kern -0.1em V\!/}c^2}}
\newcommand{\MeV}{\ensuremath{\mathrm{Me\kern -0.1em V}}}
\newcommand{\MeVc}{\ensuremath{\mathrm{Me\kern -0.1em V\!/}c}}
\newcommand{\MeVcc}{\ensuremath{\mathrm{Me\kern -0.1em V\!/}c^2}}

\newcommand{\myto}{\kern -0.3em\to\kern -0.2em}

\newcommand{\cdfii}{CDF\,II~}

\newcommand{\hh}{D^0\myto h^- h^+}
\newcommand{\kk}{D^0\myto K^- K^+}
\newcommand{\kpi}{D^0\myto K^- \pi^+}
\newcommand{\kpipio}{D^0\myto K^- \pi^+ \pi^0}

\newcommand{\pipi}{D^0\myto \pi^-\pi^+}
\newcommand{\Do}{D^0}
\newcommand{\Dst}{D^{\ast +}}

\newcommand{\KKtokpi}{\Gamma(\kk)/\Gamma(\kpi)}

\newcommand{\PIpitokpi}{\Gamma(\pipi)/\Gamma(\kpi)}
\newcommand{\KKtopipi}{\Gamma(\kk)/\Gamma(\pipi)}

\begin{document}

\title{Measurement of Partial Widths and Search for Direct CP Violation in
$D^0$ Meson Decays to $K^-K^+$ and $\pi^-\pi^+$} %

\affiliation{Institute of Physics, Academia Sinica, Taipei, Taiwan 11529, Republic of China }
\affiliation{Argonne National Laboratory, Argonne, Illinois 60439 }
\affiliation{Institut de Fisica d'Altes Energies, Universitat Autonoma de Barcelona, E-08193, Bellaterra (Barcelona), Spain }
\affiliation{Istituto Nazionale di Fisica Nucleare, University of Bologna, I-40127 Bologna, Italy }
\affiliation{Brandeis University, Waltham, Massachusetts 02254 }
\affiliation{University of California at Davis, Davis, California 95616 }
\affiliation{University of California at Los Angeles, Los Angeles, California 90024 }
\affiliation{University of California at San Diego, La Jolla, California 92093 }
\affiliation{University of California at Santa Barbara, Santa Barbara, California 93106 }
\affiliation{Instituto de Fisica de Cantabria, CSIC-University of Cantabria, 39005 Santander, Spain }
\affiliation{Carnegie Mellon University, Pittsburgh, PA 15213 }
\affiliation{Enrico Fermi Institute, University of Chicago, Chicago, Illinois 60637 }
\affiliation{Joint Institute for Nuclear Research, RU-141980 Dubna, Russia }
\affiliation{Duke University, Durham, North Carolina 27708 }
\affiliation{Fermi National Accelerator Laboratory, Batavia, Illinois 60510 }
\affiliation{University of Florida, Gainesville, Florida 32611 }
\affiliation{Laboratori Nazionali di Frascati, Istituto Nazionale di Fisica Nucleare, I-00044 Frascati, Italy }
\affiliation{University of Geneva, CH-1211 Geneva 4, Switzerland }
\affiliation{Glasgow University, Glasgow G12 8QQ, United Kingdom }
\affiliation{Harvard University, Cambridge, Massachusetts 02138 }
\affiliation{The Helsinki Group: Helsinki Institute of Physics; and Division of High Energy Physics, Department of Physical Sciences, University of Helsinki, FIN-00044, Helsinki, Finland }
\affiliation{Hiroshima University, Higashi-Hiroshima 724, Japan }
\affiliation{University of Illinois, Urbana, Illinois 61801 }
\affiliation{The Johns Hopkins University, Baltimore, Maryland 21218 }
\affiliation{Institut f\"ur Experimentelle Kernphysik, Universit\"at Karlsruhe, 76128 Karlsruhe, Germany }
\affiliation{High Energy Accelerator Research Organization (KEK), Tsukuba, Ibaraki 305, Japan }
\affiliation{Center for High Energy Physics: Kyungpook National University, Taegu 702-701; Seoul National University, Seoul 151-742; and SungKyunKwan University, Suwon 440-746; Korea }
\affiliation{Ernest Orlando Lawrence Berkeley National Laboratory, Berkeley, California 94720 }
\affiliation{University of Liverpool, Liverpool L69 7ZE, United Kingdom }
\affiliation{University College London, London WC1E 6BT, United Kingdom }
\affiliation{Massachusetts Institute of Technology, Cambridge, Massachusetts 02139 }
\affiliation{Institute of Particle Physics, McGill University, Montr\'eal, Canada H3A~2T8; and University of Toronto, Toronto, Canada M5S~1A7 }
\affiliation{University of Michigan, Ann Arbor, Michigan 48109 }
\affiliation{Michigan State University, East Lansing, Michigan 48824 }
\affiliation{Institution for Theoretical and Experimental Physics, ITEP, Moscow 117259, Russia }
\affiliation{University of New Mexico, Albuquerque, New Mexico 87131 }
\affiliation{Northwestern University, Evanston, Illinois 60208 }
\affiliation{The Ohio State University, Columbus, Ohio 43210 }
\affiliation{Okayama University, Okayama 700-8530, Japan }
\affiliation{Osaka City University, Osaka 588, Japan }
\affiliation{University of Oxford, Oxford OX1 3RH, United Kingdom }
\affiliation{University of Padova, Istituto Nazionale di Fisica Nucleare, Sezione di Padova-Trento, I-35131 Padova, Italy }
\affiliation{University of Pennsylvania, Philadelphia, Pennsylvania 19104 }
\affiliation{Istituto Nazionale di Fisica Nucleare, University and Scuola Normale Superiore of Pisa, I-56100 Pisa, Italy }
\affiliation{University of Pittsburgh, Pittsburgh, Pennsylvania 15260 }
\affiliation{Purdue University, West Lafayette, Indiana 47907 }
\affiliation{University of Rochester, Rochester, New York 14627 }
\affiliation{The Rockefeller University, New York, New York 10021 }
\affiliation{Istituto Nazionale di Fisica Nucleare, Sezione di Roma 1, University di Roma ``La Sapienza," I-00185 Roma, Italy }
\affiliation{Rutgers University, Piscataway, New Jersey 08855 }
\affiliation{Texas A\&M University, College Station, Texas 77843 }
\affiliation{Texas Tech University, Lubbock, Texas 79409 }
\affiliation{Istituto Nazionale di Fisica Nucleare, University of Trieste/\ Udine, Italy }
\affiliation{University of Tsukuba, Tsukuba, Ibaraki 305, Japan }
\affiliation{Tufts University, Medford, Massachusetts 02155 }
\affiliation{Waseda University, Tokyo 169, Japan }
\affiliation{Wayne State University, Detroit, Michigan 48201 }
\affiliation{University of Wisconsin, Madison, Wisconsin 53706 }
\affiliation{Yale University, New Haven, Connecticut 06520 }


\author{D.~Acosta}
\affiliation{University of Florida, Gainesville, Florida 32611 }

\author{T.~Affolder}
\affiliation{University of California at Santa Barbara, Santa Barbara, California 93106 }

\author{T.~Akimoto}
\affiliation{University of Tsukuba, Tsukuba, Ibaraki 305, Japan }

\author{M.G.~Albrow}
\affiliation{Fermi National Accelerator Laboratory, Batavia, Illinois 60510 }

\author{D.~Ambrose}
\affiliation{University of Pennsylvania, Philadelphia, Pennsylvania 19104 }

\author{S.~Amerio}
\affiliation{University of Padova, Istituto Nazionale di Fisica Nucleare, Sezione di Padova-Trento, I-35131 Padova, Italy }

\author{D.~Amidei}
\affiliation{University of Michigan, Ann Arbor, Michigan 48109 }

\author{A.~Anastassov}
\affiliation{Rutgers University, Piscataway, New Jersey 08855 }

\author{K.~Anikeev}
\affiliation{Massachusetts Institute of Technology, Cambridge, Massachusetts 02139 }

\author{A.~Annovi}
\affiliation{Istituto Nazionale di Fisica Nucleare, University and Scuola Normale Superiore of Pisa, I-56100 Pisa, Italy }

% ...

\collaboration{CDF Collaboration}

\begin{abstract}
We present a measurement of relative partial widths and decay rate CP asymmetries in 
$K^-K^+$ and $\pi^-\pi^+$ decays of $\Do$ mesons produced in $p\overline{p}$ 
collisions at $\sqrt{s} = 1.96\,\TeV$. 
We use a sample of $2\times10^5$ $D^{\ast+}\myto D^0\pi^+$ (and charge conjugate) 
decays with the $D^0$ decaying to $K^-\pi^+$, $K^-K^+$, and $\pi^-\pi^+$,
corresponding to $123\,{\rm pb}^{-1}$ of data collected 
by the Collider Detector at Fermilab II experiment at the Fermilab Tevatron collider.
No significant direct CP violation is observed. We measure
$\KKtokpi = 0.0992 \pm 0.0011 \pm 0.0012$, 
$\PIpitokpi = 0.03594 \pm 0.00054 \pm 0.00040$,
$A_{CP}(K^-K^+) = (2.0 \pm 1.2 \pm 0.6)\,\%$, and
$A_{CP}(\pi^-\pi^+) = (1.0 \pm 1.3 \pm 0.6)\,\%$,
where, in all cases, the first uncertainty is statistical and the second 
is systematic.
\end{abstract}


\maketitle
\begin{figure}
\includegraphics[width=8.cm,height=4.8cm]{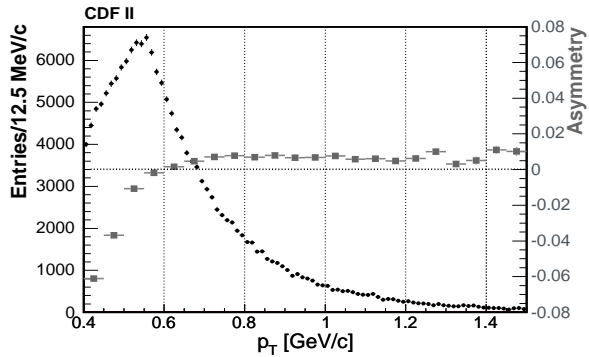}
\caption{\label{fig_qa}
The $\Dst$ decay pion $p_T$ distribution (black dots), and 
the detector charge asymmetry (gray squares) as a function of track 
$p_T$.}
\end{figure}

The Cabibbo suppressed decays $D^0 \myto K^-K^+, \pi^-\pi^+$ have been used to 
study $D^0$ mixing and CP violation in the charm sector.
Direct CP violation in decay rates requires the interference of two amplitudes 
with different weak and strong phases.
In $D^0\myto K^-K^+, \pi^-\pi^+$, the spectator and penguin amplitudes have 
different weak phases, and different strong phases
are expected to be generated by rescattering in final state interactions (FSI).
The predicted rates of CP violation are of the order 
of the imaginary part of the $V_{cs}$ element of the CKM matrix,
$O(0.1\%)$. New physics, providing additional phases, can 
enhance these predictions up to $O(1\%)$~\cite{predict}.
At present there is no experimental evidence of direct CP violation in these 
decays; a combination of previous measurements~\cite{pdg2004} yields,
for the direct CP asymmetries ($A_{CP}$),
$A_{CP}(K^-K^+) = 0.005 \pm 0.016$, and $A_{CP}(\pi^-\pi^+) = 0.021 \pm 0.026$.
\newline
\indent
In the limit of exact SU(3) flavor symmetry~\cite{SU3} $\KKtopipi \sim 1$.
Including the effects of phase space, the difference of the kaon and pion decay 
constants and other SU(3) breaking effects
may increase this ratio up to 1.4~\cite{buras}.
The world average value is $2.826 \pm 0.097$~\cite{pdg2004},
well above the expectations. Large FSI and 
contributions from penguin diagrams have been proposed to explain 
this discrepancy~\cite{BRpuzzle}. 
Phenomenological analyses~\cite{buccella}, using available data on 
$\Do$ and $D^+$ branching ratios,  
derive the magnitudes and phase shifts of the relevant amplitudes, including FSI, 
that reproduce the above world-average measured ratio.
The same phenomenological analyses predict CP asymmetries as high
as $0.1\%$ for certain Cabbibo-suppressed decays and somewhat lower
asymmetries for the $K^-K^+$ and $\pi^-\pi^+$ channels.
A significant asymmetry at the level of $1\%$, not yet excluded experimentally,
would be an interesting indication for nonstandard model sources of CP violation 
in the charm sector.
\newline
\indent We present measurements
of the ratios $\KKtokpi$, and $\PIpitokpi$, and results of the search for 
direct CP violation in the Cabibbo-suppressed $\kk$ and $\pipi$ decays.
The sample contains 
$2\kern-0.1em \times\kern-0.1em 10^5$ $\Dst \myto \Do \pi^+$ events, 
with $\Do$ decaying to the three modes under study
(charge conjugate states are implied 
throughout this paper, unless otherwise stated).
The $\Do$ flavor is unambiguously determined from the charge of the 
pion in the strong decay  $\Dst \myto \Do \pi^+$.
\newline 
\indent The components of
the \cdfii detector pertinent to this analysis are described briefly
below; a more complete description can be found elsewhere~\cite{cdf}. 
For this measurement we use only 
tracks reconstructed by both the Central Outer Tracker (COT)~\cite{cot} and
the silicon microstrip detector (SVX\,II)~\cite{svxii} in the pseudorapidity 
range $|\eta|\lesssim 1$~\cite{CDFsystem}.
The $D^0$ decays used in this analysis are selected with a 
three-level trigger system. 
At Level~1, charged tracks are reconstructed
in the COT transverse plane by a hardware processor (eXtremely Fast Tracker)~\cite{xft}.
The trigger requires two oppositely charged tracks with transverse
momenta $p_T\ge2\,\GeVc$ and the scalar sum 
$p_{T1}+p_{T2}\ge5.5\,\GeVc$. At 
Level~2, the Silicon Vertex Tracker (SVT)~\cite{svt} associates SVX\,II
$r$-$\phi$ position measurements with XFT tracks, providing 
a precise measurement of the track impact parameter ($d_0$), 
defined as the distance of closest approach, in the transverse plane, of 
the trajectory of the track to the beam axis.
The resolution of this impact parameter measurement is $50\,\mu\rm m$, which 
includes a $\approx 30\rm\,\mu m$ contribution from the 
transverse beam size.
Hadronic decays of heavy flavor particles are selected by requiring 
two-tracks (trigger tracks) with $120\,\mu{\rm m}\le d_0 \le1.0\rm\,mm$.
The two trigger tracks must have an opening angle 
in the transverse plane satisfying $2^\circ\le|\Delta\phi|\le90^\circ$
and must satisfy the requirement $L_{\rm xy}>200\rm\,\mu m$,
where the two-dimensional decay length, $L_{\rm xy}$, is calculated
as the transverse distance from the beam line to the two-track 
vertex projected along the total transverse momentum of the 
track pair.
At Level~3, a complete event reconstruction is performed, 
and the Level~1 and Level~2 requirements are confirmed.
\newline
\indent The reconstruction of $\Dst$ candidates starts from the selection 
of pairs of oppositely charged tracks that satisfy the trigger requirements.
We form one $\kpi$, $K^-K^+$, and $\pi^-\pi^+$ 
candidate for each trigger pair. For the $K^-\pi^+$ mode we also form a second 
$\Do$ candidate with the mass assignments interchanged. No $K$ or $\pi$ 
particle identification is used in this analysis.
$\Do$ candidates whose invariant mass is within $\pm 100\,\MeVcc$ 
of the mean reconstructed $\Do$ mass
are combined with a third track
with $p_T \ge 0.4\,\GeVc$ to form a $D^{\ast +}\myto\Do\pi^+$ candidate. 
In the reconstruction of $\kpi$ decays, the charge of the pion
from the $\Do$ decay is required to be the same as the charge of the pion
from the $\Dst$ decay.
\newline
\indent To reduce combinatorial background and background from 
partially reconstructed $D^0$ decays, we require the measured 
mass difference, $\Delta M$, between the $\Dst$ and $\Do$ mesons 
to be within three standard deviations in experimental resolution 
of the expected value:
$143.5\,\MeVcc < \Delta M < 147.2\,\MeVcc$. 
Finally, to reduce the potential systematic uncertainty induced by the different 
acceptance ratios of $\Dst$ produced in $B$-hadron decays, the 
contribution ($\sim\kern -0.3em 12\%$)~\cite{xscharm} 
of non-prompt $\Dst$ is reduced by requiring the impact parameter of 
the $\Do$ meson to satisfy $d_0(D^0) \le 100\,\mu {\rm m}$.
\begin{figure}
\includegraphics[width=9.4cm,height=4.7cm]{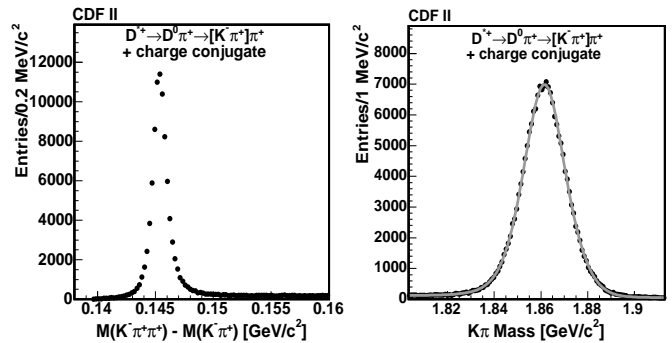}
\caption{\label{fig_kpi}
The $\Delta M = M(K^-\pi^+\pi^+) - M(K^-\pi^+)$ distribution (left) for the
$D^0\myto K^-\pi^+$ candidates. The $K^-\pi^+$ invariant mass distribution 
(right) after all selection criteria 
have been applied. The curve is the sum of the fits performed separately 
for the $D^0$ and $\overline{D^0}$ mesons.}
\end{figure}
\begin{figure}
\includegraphics[width=8.5cm,height=4.6cm]{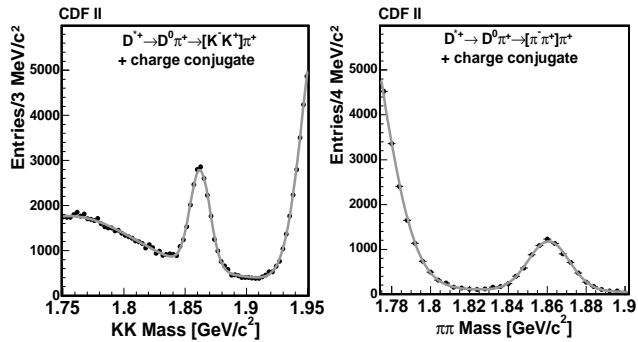}
\caption{\label{fig_2pi_2k}
The $K^-K^+$ (left) and $\pi^-\pi^+$ (right) invariant mass distributions
after all selection criteria have been applied.}
\end{figure}
\newline
\indent The $D^0$ yields are obtained from binned maximum likelihood fits 
to the $\Do$ invariant mass distributions.
For the $K^-\pi^+$ mode, the signal is modeled with a single Gaussian function plus a 
convolution of an exponential function with an error function to 
model the low mass tail of the observed distribution;
a second-degree polynomial is used to model 
the combinatorial background. 
For the $K^-K^+$ and $\pi^-\pi^+$ modes, due to the limited event statistics, 
we use a single Gaussian as a model for the signal.
We use Gaussian functions to describe 
both the $K^-\pi^+$ mis-identification peaks in the $K^-K^+$ and 
$\pi^-\pi^+$ modes and the background from partially reconstructed 
$\kpipio$ decays in the $K^-K^+$ mode, and we verified,
in simulated samples of inclusive $D^0$ decays, that this model 
adequately describes both sources of background.
The invariant mass distributions for the $K^-\pi^+$, $K^-K^+$, and $\pi^-\pi^+$ 
modes are shown in Fig.~\ref{fig_kpi} and Fig.~\ref{fig_2pi_2k}.
The number of signal events from the fits to the invariant mass 
distributions are reported in Table~\ref{tab:events}.
\begin{table}[h]
\caption{\label{tab:events} The $D^0$ and $\overline{D^0}$ signals determined 
from the fits to the invariant mass distributions. The errors are the statistical 
uncertainties from the fits.}
\begin{ruledtabular}
\begin{tabular}{lccc}
Mode     &   $D^0    $ & $\overline{D^0}$ & Total \\\hline
$K\pi$   & $88,310 \pm 330$ & $92,600 \pm 340$ & $180,910 \pm 480$ \\
$KK$     & $8,190 \pm 140$  & $8,030 \pm 140$   & $16,220 \pm 200$ \\
$\pi\pi$ & $3,660 \pm 69$   & $3,674 \pm 68$   & $7,334 \pm 97$ \\
\end{tabular}
\end{ruledtabular}
\end{table}
\newline
\indent The relative branching fractions are extracted using the formula
\begin{equation}
\label{eq:BRhh}
\frac{\Gamma\left(\hh \right)}{\Gamma\left( \kpi \right)} =
\frac{N_{h^-h^+}}{N_{K\pi}} \cdot \frac{\epsilon_{K\pi}}{\epsilon_{h^-h^+}} =
\frac{N_{h^-h^+}}{N_{K\pi}} \cdot R_{h^-h^+},
\end{equation}
\noindent where $h=K$ or $\pi$, $N_{h^- h^+}$ is the total number of $\Do$ mesons decaying in the 
appropriate mode from Table~\ref{tab:events}, and $\epsilon_{h^- h^+}$ is the 
average $\Do$ and $\overline{\Do}$ acceptance for each of the decays, including 
trigger and reconstruction efficiency. The quantity $R_{h^-h^+}$ is the efficiency 
ratio of the $\kpi$ to $\hh$ mode.
\newline
\indent We have used a Monte Carlo simulation, based on GEANT~\cite{geant}, of the 
\cdfii detector and trigger to determine the ratios of the relative trigger and 
reconstruction efficiencies for the three decay modes.
The trigger efficiency varies among the three modes due to the different 
nuclear interaction and decay-in-flight probabilities for $\pi^+$, $\pi^-$, $K^+$, and $K^-$,
the differences in the kinematics of the decay ({\it e.g.}, opening angle distributions),
induced by the masses of the final state particles, and the different 
XFT efficiency as a function of the track $p_T$ caused by the different specific 
ionization in the COT for  $\pi^\pm$ and $K^\pm$.
The simulated signals have been generated using as input the momentum and rapidity 
distributions of the $\Dst$ mesons as measured by \cdfii\cite{xscharm}.
The simulation of the \cdfii detector includes the time variation
of the beam position and of the hardware configuration in
the SVX\,II and SVT.
The trigger efficiencies have been studied in detail using calibration 
samples of real data.
For the ratio of efficiencies we obtain
$R_{KK} = 1.1073 \pm 0.0074$ and $R_{\pi\pi} = 0.8867 \pm 0.0056$,
where the uncertainties are due to Monte Carlo statistics.
For the relative $\kk$ to $\pipi$ efficiencies we obtain $1.2488 \pm 0.0078$.
\newline
\indent The systematic uncertainty on the ratios of the signal yields due to 
the fitting procedure has been estimated by varying the model used for 
the combinatorial background 
(using a third-degree polynomial instead of a second-degree polynomial), using two 
Gaussian functions with different means and widths to describe $\Do$ signals,
and performing the fits in different ranges of $p_T(D^0)$.
This systematic uncertainty is listed in the first row of Table~\ref{tab:systbr}.
We have evaluated the systematic uncertainty in the determination of 
the relative efficiencies from the following sources: Monte Carlo statistic,
the simulation of the XFT and SVT triggers, the time-dependent variations of the beam spot 
size in $z$, the simulation of nuclear interactions in the \cdfii detector, 
the effect on the trigger efficiency due to a possible lifetime difference
between the CP-even and CP-mixed $\Do$ decays, the input $p_T$ spectra for $\Dst$ mesons, and 
the different ratios of efficiencies for $\Dst$ produced in $B$-hadron decays. 
The contribution of each source listed above to the
total relative systematic error on the ratio of branching fraction measurements 
is reported in Table~\ref{tab:systbr}.
\begin{table}[h]
\caption{\label{tab:systbr}
The sources of systematic uncertainty on the ratios
of branching fractions and their contributions to the total fractional
systematic uncertainty.} 
\begin{ruledtabular}
\begin{tabular}{lccc}
Systematic source         & $(\frac{KK}{K\pi})$[\%] &  $(\frac{\pi\pi}{K\pi})$[\%] & $(\frac{KK}{\pi\pi})$[\%]\vspace*{0.5mm} \\\hline
Signal Yields             & $0.64$ & $0.54$  & $0.67$ \\
Monte Carlo statistics    & $0.67$ & $0.63$  & $0.62$ \\
Trigger simulation        & $0.34$ & $0.31$  & $0.37$ \\
Beam spot size            & $0.35$ & $0.24$  & $0.35$ \\
Material in GEANT         & $0.28$ & $0.30$  & $0.59$ \\
Lifetime difference       & $0.55$ & $0.55$  &  \\
Input spectra             & $0.05$ & $<0.01$ & $<0.01$  \\
Non prompt $D^\ast$       & $0.16$ & $0.08$  & $0.24$ \\\hline
Total relative error:     & $1.2 $ & $1.1 $  & $1.2 $  \\
\end{tabular}
\end{ruledtabular}
\end{table}
%
Using Eq.~\ref{eq:BRhh} we derive the relative branching 
ratios reported in Table~\ref{tab:summResults}.
In addition, we derive $\KKtopipi = 2.760 \pm 0.040\,(stat) \pm 0.034\,(syst)$.
\newline
\indent We extract the CP decay rate asymmetries, using the same samples 
of $\Do$ decays described above, by measuring
\begin{eqnarray*}
A_{CP} \equiv \frac{\Gamma ( D^0
\myto f ) - \Gamma ( \overline{D^0} \myto f )}{
\Gamma ( D^0 \myto f ) + \Gamma ( \overline{D^0}
\myto f )},
\end{eqnarray*}
\noindent where $f$ represents either the $K^- K^+$ or $\pi^- \pi^+$ final 
state.
The direct production of charm mesons in $p\overline{p}$ collisions is assumed
to be CP invariant. The measured CP asymmetry must be
corrected for different detector efficiencies (detector charge asymmetry)
for positive and negative charged pions in the $D^\ast$ decay, 
which produce a different 
detection efficiency for $D^{\ast +}$ and $D^{\ast -}$ mesons.
\newline
\indent The detector charge asymmetry is produced by the interactions
of particles with the detector material and by effects related to the 
cell geometry of the COT.
We measure this asymmetry in order to correct
the number of observed
$D^{\ast +} \myto D^0 \pi^+$ decays
relative to the number of observed
$D^{\ast -} \myto \overline{D^0} \pi^-$ decays
for the difference in detection efficiencies of $\pi^+$ and $\pi^-$.
For the detector charge asymmetry measurement,
we compare the numbers of reconstructed positive and negative tracks 
as a function of track $p_T$ in a high statistics data sample
collected with the same trigger used to collect the signal sample.
We avoid a bias in the charge asymmetry due to interactions of
the beam with material in the detector near the interaction region
by selecting tracks which originate from the primary $p\overline{p}$ 
collision point, requiring the track impact parameter
to be $d_0 \le100\rm\,\mu m$.
The detector charge asymmetry, defined as
$(N^+ - N^-)/(N^+ + N^-)$, where $N^{+}$ 
($N^{-}$) 
is the number of positive (negative) tracks in the sample,
is shown as a function of the track $p_T$ in Fig.~\ref{fig_qa}.
Using the event yields in Table~\ref{tab:events}, and correcting
for the detector charge asymmetry, we obtain the CP asymmetries
reported in Table~\ref{tab:summResults}.
\begin{table}[h]
\caption{\label{tab:summResults} 
Summary of results from this analysis. The first uncertainty is statistical,
the second systematic.}
\begin{ruledtabular}
\begin{tabular}{lcc}
                          & $\kk ~[\%] $  &  $\pipi ~[\%]$ \\\hline
$\Gamma/\Gamma(K^-\pi^+)$ & $ 9.92 \pm 0.11\,\pm 0.12$ 
                          & $ 3.594 \pm 0.054\,\pm 0.040$ \\
$A_{CP}$                  & $2.0 \pm 1.2\,\pm 0.6$ 
                          & $ 1.0 \pm 1.3\,\pm 0.6$  \\
\end{tabular}
\end{ruledtabular}
\end{table}
\newline
\indent To evaluate the systematic uncertainty associated with the charge
asymmetry corrections we apply the corrections to the sample of
$D^{\ast +} \myto D^0 \pi^+ \myto [K^-\pi^+]\pi^+$
decays, where, in the standard model, we expect no 
CP violation.
Unlike the analysis for the decays to CP eigenstates, in this case
we must also apply an efficiency correction of $3\%$ due to the different
nuclear interaction rates of $K^+$ and $K^-$, derived from 
the Monte Carlo described above.
A residual asymmetry of $(0.35 \pm 0.53)\%$ is found, where the error 
is the statistical uncertainty due to the data and Monte Carlo statistics.
In addition, we check possible dependence of the charge asymmetry corrections
on the event environment by deriving the corrections
using track samples selected by different triggers
and using a sample of $K_S^0 \myto \pi^-\pi^+$ decays instead of generic 
tracks.
We also check for charge dependent effects on the observables
used in the analysis ($\Delta M$ and $D^0$ invariant mass) and 
in the signal shapes. In all cases we find negligible effects.
Finally we test the quality of the charge asymmetry corrections 
by performing the CP asymmetry measurements 
dividing the signal samples into two ranges of
$\Dst$ pion transverse momentum ($p_{T}> 0.6\,\GeVc$ and 
$p_{T}\le 0.6\,\GeVc$).
These additional uncertainty estimates result in variations smaller than the
uncertainty of $\pm 0.53\%$ on the asymmetry measurement 
described above, and this statistical uncertainty is adopted as a conservative 
estimate of our systematic error.
An additional systematic uncertainty of $\pm 0.2\%$, due to the yield determination
of $\Do$ and $\overline{\Do}$, is added in quadrature
to the detector charge asymmetry correction uncertainty;
other sources give negligible contributions and are ignored.
\newline
\indent In summary, we have used the \cdfii detector to measure
the ratios of partial widths
$\KKtokpi = 0.0992 \pm 0.0011\,(stat) \pm 0.0012\,(syst)$,
$\PIpitokpi = 0.03594 \pm 0.00054\,(stat) \pm 0.00040\,(syst)$.
These measurements agree with, and are an improvement in
precision over, the world averages
$\KKtokpi = 0.1023^{+0.0022}_{-0.0027}$, 
$\PIpitokpi = 0.0362 \pm 0.0010$~\cite{pdg2004}.
\indent We have made the most precise measurement to date of the direct CP asymmetries
$A_{CP}(K^-K^+)=[2.0 \pm 1.2\,(stat) \pm 0.6\,(syst)]\%$, and
$A_{CP}(\pi^-\pi^+)=[1.0 \pm 1.3\,(stat) \pm 0.6\,(syst)]\%$.
In agreement with the world averages
$A_{CP}(K^-K^+)=(0.5 \pm 1.6)\%$, and
$A_{CP}(\pi^-\pi^+)=(2.1 \pm 2.6)\%$~\cite{pdg2004}.
At present there is no evidence for direct CP violation in 
Cabibbo-suppressed $\Do$ decays. 

We thank the Fermilab staff and the technical staffs of the participating institutions 
for their vital contributions. This work was supported by the U.S. Department of Energy 
and National Science Foundation; the Italian Istituto Nazionale di Fisica Nucleare; 
the Ministry of Education, Culture, Sports, Science and Technology of Japan; 
the Natural Sciences and Engineering Research Council of Canada; 
the National Science Council of the Republic of China; the Swiss National Science Foundation; 
the A.P. Sloan Foundation; the Bundesministerium fuer Bildung und Forschung, Germany; 
the Korean Science and Engineering Foundation and the Korean Research Foundation; 
the Particle Physics and Astronomy Research Council and the Royal Society, UK; 
the Russian Foundation for Basic Research; 
the Comision Intermi\-nis\-terial de Ciencia y Tecnologia, Spain; 
and in part by the European Community's Human Potential Programme under 
contract HPRN-CT-2002-00292, Probe for New Physics.


\begin{thebibliography}{99}
\bibitem{CDFsystem}
  \cdfii uses a cylindrical coordinate system in which $\phi$ is the azimuthal
angle, $r$ is the radius from the nominal beamline, $y$ points up and $z$
points in the proton beam direction with the origin at the center of the detector.
The transverse plane is the plane perpendicular to the $z$ axis.
A superconducting magnet provides a nearly uniform $1.4\,{\rm T}$ axial field
in which charged particles are reconstructed.
\end{thebibliography}
\end{document}